\documentclass[12pt]{article} 
\usepackage{cite}
\usepackage[utf8]{inputenc}
\usepackage{amsmath}

\usepackage{graphicx}

\begin{document}
\title{Reproducing entanglement through local classical resources with no communication}
\author{Antonio Di Lorenzo\\ 
Instituto de F\'{\i}sica, Universidade Federal de Uberl\^{a}ndia, \\
38400-902 Uberl\^{a}ndia, Minas Gerais, Brazil}
\date{}
\maketitle
Entanglement is one of the most intriguing features of quantum mechanics. 
It gives rise to peculiar correlations which cannot be reproduced by a large class of alternative 
theories, the so-called hidden-variable models, that use parameters in addition to the wave-function. 
This incompatibility was quantified through the celebrated Bell inequalities \cite{Bell1964,Clauser1969,Clauser1974}, 
and more recently through new inequalities due to Leggett \cite{Leggett2003,Groblacher2007,Branciard2008}. 
Experiments confirm the predictions of quantum mechanics \cite{Freedman1972,Aspect1982,Tapster1994,Weihs1998,Tittel1998,Rowe2001,
Groblacher2007a,Paterek2007,Branciard2007,Branciard2008,Eisaman2008,Romero2010,Paternostro2010}. 
However, this does not imply that quantum mechanics is the ultimate theory, unsusceptible of improvement, nor that quantum mechanics is essentially non-local. The theories ruled out by Bell and Leggett inequalities are required 
to satisfy some hypotheses, none of which is implied by locality alone. 
By dropping one or more hypotheses, it is possible not only to violate said inequalities, 
but to reproduce the quantum mechanical predictions altogether \cite{Groblacher2007,Brans1988,Toner2003,DeZela2008,Hall2010,Hall2011,DiLorenzo2011c,DiLorenzo2011e}. 
So far, the models proposed were only mathematical constructs. In this paper we provide a classical 
realization of two \cite{Hall2010,DiLorenzo2011c} of these models, using local classical resources, without
 recurring to any type of communication among the involved parties. 
The resources consist in two baseballs, two bats, and a number of synchronized watches. 
Our results demonstrate the possibility of reproducing the quantum mechanical correlations, and even creating 
stronger correlations which provide the maximum violation of the Bell inequality, beyond the Cirel'son bound 
\cite{Cirelson1980}
for quantum mechanics. 

\begin{figure}
\includegraphics[width=4in]{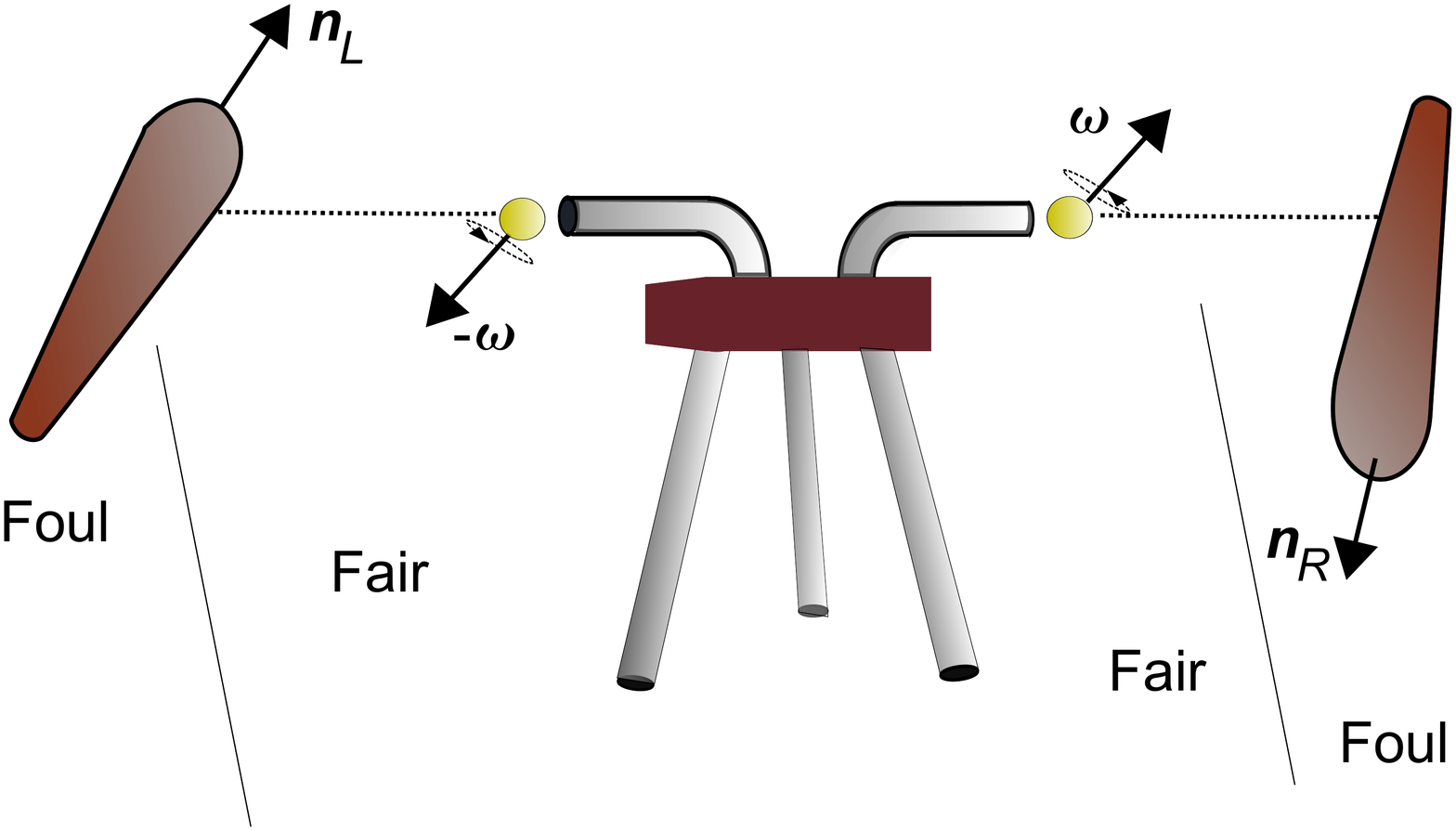}
\caption{\label{fig:setup} Scheme of the setup.}
\end{figure}

\begin{figure}
\includegraphics[width=4in]{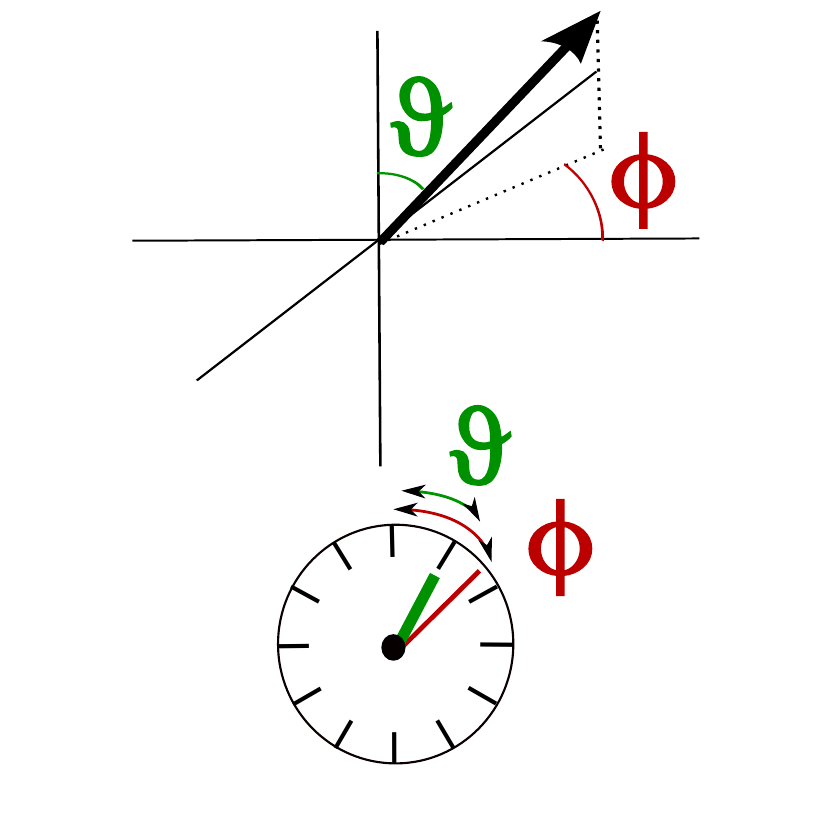}
\caption{\label{fig:watch} A unit vector is associated with the hands of a watch.}
\end{figure}

Let us consider the following setup depicted in Fig. \ref{fig:setup}: 
there is a sophisticated baseball pitching machine able to pitch two balls spinning with angular velocities 
 $\boldsymbol{\omega}$  and $-\boldsymbol{\omega}$ (with $\omega$ fixed)
in opposite directions. The balls are approximatedly a rigid body and have spherical symmetry. 
They are pitched with fixed center-of-mass velocities in a vacuum (so that the spin does not 
influence the center-of-mass trajectory through the Magnus effect) 
such that their centers of mass follow given trajectories each ending 
in a bat controlled by an independent system. 
Each bat is carefully crafted to be a solid of revolution, its center of mass is being held fixed, 
and a machine varies the orientation $\mathbf{n}$ of the bat's symmetry axis.    
It is empirically found that a ball pitched with $\boldsymbol{\omega}=\omega \mathbf{u}$, 
after hitting a bat whose axis is oriented along $\mathbf{n}$, will fall in the foul ground with 
a frequency $(1-\mathbf{n}\cdot\mathbf{u})/2$, and in the fair ground with the complementary 
frequency $(1+\mathbf{n}\cdot\mathbf{u})/2$. 
Due to our politically incorrect bias against negative numbers, 
we shall associate the value $\sigma=-1$ to the event of the ball being batted in the foul ground, 
and $\sigma=+1$ to the alternative event. 
The pitching machine varies the spin of the two balls according to the following algorithm: 
it flips a fair coin, and according to the result, heads $H$ or tails $T$, 
it consults either of two internal watches, $W_H$ and $W_T$. 
Each watch is built so that the small hand has a period $\tau_{w,s}$ and the large hand 
$\tau_{w,l}$, with $w\in\{H,T\}$. Unlike what happens in ordinary watches, 
any two of the four periods are mutually incommensurable.
The positions of the hands are used to determine a unit vector $\mathbf{n}$. 
This may happen in a straightforward fashion, as depitced in Fig. \ref{fig:watch} 
or by using a Montecarlo algorithm, 
so that an external observer could not predict what spin is chosen, even knowing 
the periods of the watches. 
The machine will then pitch a ball spinning along $\mathbf{n}$ to the left, and one 
spinning along $-\mathbf{n}$ to the right, or \emph{viceversa}, according to another 
coin flip, with outcome $d\in\{H,T\}$. 

The batting machine to the right possesses a watch $W'_H$, and the one to the left a watch 
$W'_T$, which have the same periods as the watches of the pitching machine, 
but are engineered to rotate counterclockwise, and are synchronized so that 
each hand of $W'_w$ passes through the conventional zero position at the same time the 
corresponding hand of $W_w$ does. 
Thus the sum of the times indicated by the 
hand $h$, $h\in\{s,l\}$, of the watches $W_w$ and $W'_w$
is zero modulo $\tau_{w,h}$, i.e., we have four conserved quantities. 
Each batting machine is using the same algorithm as the pitching machine to determine 
the value of $\mathbf{n}_w$, but it is subtracting 
from the input values $t_{w,h}$ the number $\Delta t \ (\mathrm{mod}\ \tau_{w,h})$, 
where $\Delta t$ is the time-of-flight of the ball.

We remind that the probability of the outcome $\sigma$
for a ball spinning in the direction $\mathbf{u}$ and hitting a bat oriented along $\mathbf{n}$ 
is 
\begin{equation}
P_{\sigma}(\mathbf{n},\mathbf{u})=\frac{1}{2}\left[1+\sigma\mathbf{n}\cdot\mathbf{u}\right] . 
\end{equation}
It is easy to check that the expected joint frequency of observing the outcomes 
$\sigma,\tau$ to the left and to the right when the respective bats have 
orientations $\mathbf{n}_L,\mathbf{n}_R$ (still guided by politically incorrect bias, 
we identify the indices $T\equiv L\equiv -1$, $H\equiv R\equiv 1$)
is 
\begin{align}
\nonumber
P_{\sigma,\tau}(\mathbf{n}_L,\mathbf{n}_R)&
=\int\! d\mathbf{u}d\mathbf{v}\ \rho(\mathbf{u},\mathbf{v}|\mathbf{n}_L,\mathbf{n}_R)
P_{\sigma}(\mathbf{n}_L,\mathbf{u})
P_{\tau}(\mathbf{n}_R,\mathbf{v})\\
&=
\frac{1}{4}\left[1-\sigma\tau\mathbf{n}_L\cdot\mathbf{n}_R\right] ,
\end{align}
with the density 
\begin{equation}\label{eq:dilorconddistr}
\rho(\mathbf{u},\mathbf{v}|\mathbf{n}_L,\mathbf{n}_R)=
\frac{1}{4}\delta(\mathbf{u}+\mathbf{v})
\sum_{w,d}\delta(\mathbf{u}-d\,\mathbf{n}_w) \\
\end{equation}

This model constitutes the realization, through local classical resources, of the construction proposed in Ref.~\cite{DiLorenzo2011c}. While in the cited paper the model was presented as non-local, since it can be realized 
also through a hypothetical non-local interaction for which the fixing of the $\mathbf{n}_L,\mathbf{n}_R$ 
determines the possible values of $\mathbf{u},\mathbf{v}$, it was hinted that a local realization is possible. 
Here, we have just presented this realization. 
We notice that the batting machines are devoid of ``free will'', since the directions $\mathbf{n}_L,\mathbf{n}_R$ 
are determined, while the pitching machine has two binary choices, first in choosing either watch, 
then in choosing which ball to pitch towards, e.g., left. 
Indeed, using the ``free will'' quantifier introduced in Ref.~\cite{Hall2010}, we have that 
\begin{equation}
M=\mathrm{sup}\int d\mathbf{u}d\mathbf{v} 
\left|\rho(\mathbf{u},\mathbf{v}|\mathbf{n}_L,\mathbf{n}_R)-\rho(\mathbf{u},\mathbf{v}|\mathbf{n}'_L,\mathbf{n}'_R)\right|=2 .
\end{equation}
This is the maximum value of $M$ and corresponds to the absence of ``free will''. 

Ref.~\cite{Hall2010} proposes a different decomposition of the quantum mechanical probability, through a model 
which has the maximum ``free will'' compatible with quantum mechanics, and deterministic outcomes: 
\begin{equation}\label{eq:halldens}
\rho(\mathbf{u},\mathbf{v}|\mathbf{n}_L,\mathbf{n}_R)=\delta(\mathbf{u}+\mathbf{v}) 
\frac{1-f(\mathbf{u},\mathbf{v}, \mathbf{n}_L,\mathbf{n}_R)}{8\arccos{f(\mathbf{u},\mathbf{v}, \mathbf{n}_L,\mathbf{n}_R)}},
\end{equation}
where 
\begin{equation}
f(\mathbf{u},\mathbf{v}, \mathbf{n}_L,\mathbf{n}_R)=
\mathrm{sgn}(\mathbf{u}\cdot\mathbf{n}_L)
\mathrm{sgn}(\mathbf{v}\cdot\mathbf{n}_R)\ \mathbf{n}_L\cdot\mathbf{n}_R ,
\end{equation}
and $\mathrm{sgn}(x)=1$ for $x\ge 0$, $\mathrm{sgn}(x)=-1$ for $x< 0$.  
The probability of the outcomes, given $\mathbf{u},\mathbf{v}$, is deterministic
\begin{equation}
P_{\sigma,\tau}(\mathbf{u},\mathbf{v}, \mathbf{n}_L,\mathbf{n}_R)=
\delta_{\sigma,\mathrm{sgn}(\mathbf{u}\cdot\mathbf{n}_L)} \delta_{\tau,\mathrm{sgn}(\mathbf{v}\cdot\mathbf{n}_R)}.
\end{equation}

It is immediate to realize that the model of Ref.~\cite{Hall2010} can be reproduced by changing the algorithm used by 
the pitching machine. 
Now the pitching machine will use both watches $W_H,W_T$ 
to determine two unit vectors $\mathbf{n}_R,\mathbf{n}_L$. 
The pitcher is in possession of a third watch $W_0$, whose hands determine a unit vector $\mathbf{u}$. 
The third watch, however, is not ticking regularly, but it is coupled to the watches $W_R$ and $W_L$ in such 
a way that the hands correspond to the vector $\mathbf{u}$, for given $\mathbf{n}_L,\mathbf{n}_R$, with 
the frequency 
\begin{equation}\label{eq:halldens2}
\Pi(\mathbf{u}|\mathbf{n}_L,\mathbf{n}_R)= 
\frac{1-f(\mathbf{u},-\mathbf{u}, \mathbf{n}_L,\mathbf{n}_R)}{8\arccos{f(\mathbf{u},-\mathbf{u}, \mathbf{n}_L,\mathbf{n}_R)}}. 
\end{equation}
The pitcher will then proceed to pitch two balls, the first, spinning about $\mathbf{u}$, to the left, and the second, 
spinning about $-\mathbf{u}$, to the right. 
The other parameters of the procedure $\omega,v$ are readjusted such that a ball spinning around 
$\mathbf{u}$ hitting a bat oriented along $\mathbf{n}$ will give the outcome $\sigma$ with 
probability 
\begin{equation}\label{eq:hallcondprob}
P_\sigma(\mathbf{u},\mathbf{n})=\delta_{\sigma,\mathrm{sgn}(\mathbf{u}\cdot\mathbf{n})}.
\end{equation}
The batters, on the other hand, keep using the former algorithm in order to determine $\mathbf{n}_L,\mathbf{n}_R$. 
A problem revealed by this realization is that the ``free will'' of the batting machines is still 
determined as before. 
The model, however, admits the following alternative realization: 
the pitching machine and the two batting machines possess an identical watch $W_0$ each. 
In addition to $W_0$, the batters possess two pairs of identical watches $W_L,W_R$ each. 
The pitching machine chooses a random direction $\mathbf{u}$ according to $W_0$, as specified above, but with $W_0$ 
ticking freely now, so that the probability is uniformly distributed. 
At the moment of the pitching (or afterwards), the batting machines determine the same 
two vectors $\mathbf{n}_L$, $\mathbf{n}_R$ by looking at their watches $W_L,W_R$. 
Now, however, these watches are coupled to $W_0$ in such a way that they are influenced by it. 
For a given value of $\mathbf{u}$, the vectors $\mathbf{n}_L,\mathbf{n}_R$ appear with the 
following probability density
\begin{equation}
\Pi_2(\mathbf{n}_L,\mathbf{n}_R|\mathbf{u})= \frac{1}{4\pi}\Pi(\mathbf{u}|\mathbf{n}_L,\mathbf{n}_R) , 
\end{equation}
with $\Pi(\mathbf{u}|\mathbf{n}_L,\mathbf{n}_R)$ given in Eq.~\eqref{eq:halldens2}.
The left batting machine will pick the first vector, and the right one the second. 
In this second realization, the ``free will'' of the batters is preserved at the expense of the pitcher's. 
It seems thus that the measure $M$ introduced in Ref.~\cite{Hall2010} quantifies the maximum 
``free will'' that the receiving stations may possibly have, but does not imply that they actually have it, since 
a ``slave will'' realization of the same probability distribution is possible. 

Finally, we show that with the same classical setup it is possible to violate the Cirel'son bound, which establishes 
the maximum violation of Bell inequality that quantum mechanics can achieve ($2\sqrt{2}$, when the Bell limit is $2$). 
All we need to do is to mix the two models of Ref.~\cite{Hall2010} and \cite{DiLorenzo2011c}: 
the balls are pitched with velocities such that Eq.~\eqref{eq:hallcondprob} holds. 
The distribution of spins, however, obeys Eq.~\eqref{eq:dilorconddistr}, not Eq.~\eqref{eq:halldens}. 
This gives the joint probability 
\begin{align}\label{eq:mixprob}
P_{\sigma,\tau}(\mathbf{n}_L,\mathbf{n}_R)&
=\frac{1}{4}\left[1-\sigma\tau\mathrm{sgn}(\mathbf{n}_L\cdot\mathbf{n}_R)\right] .
\end{align}
\begin{figure}
\includegraphics[width=4in]{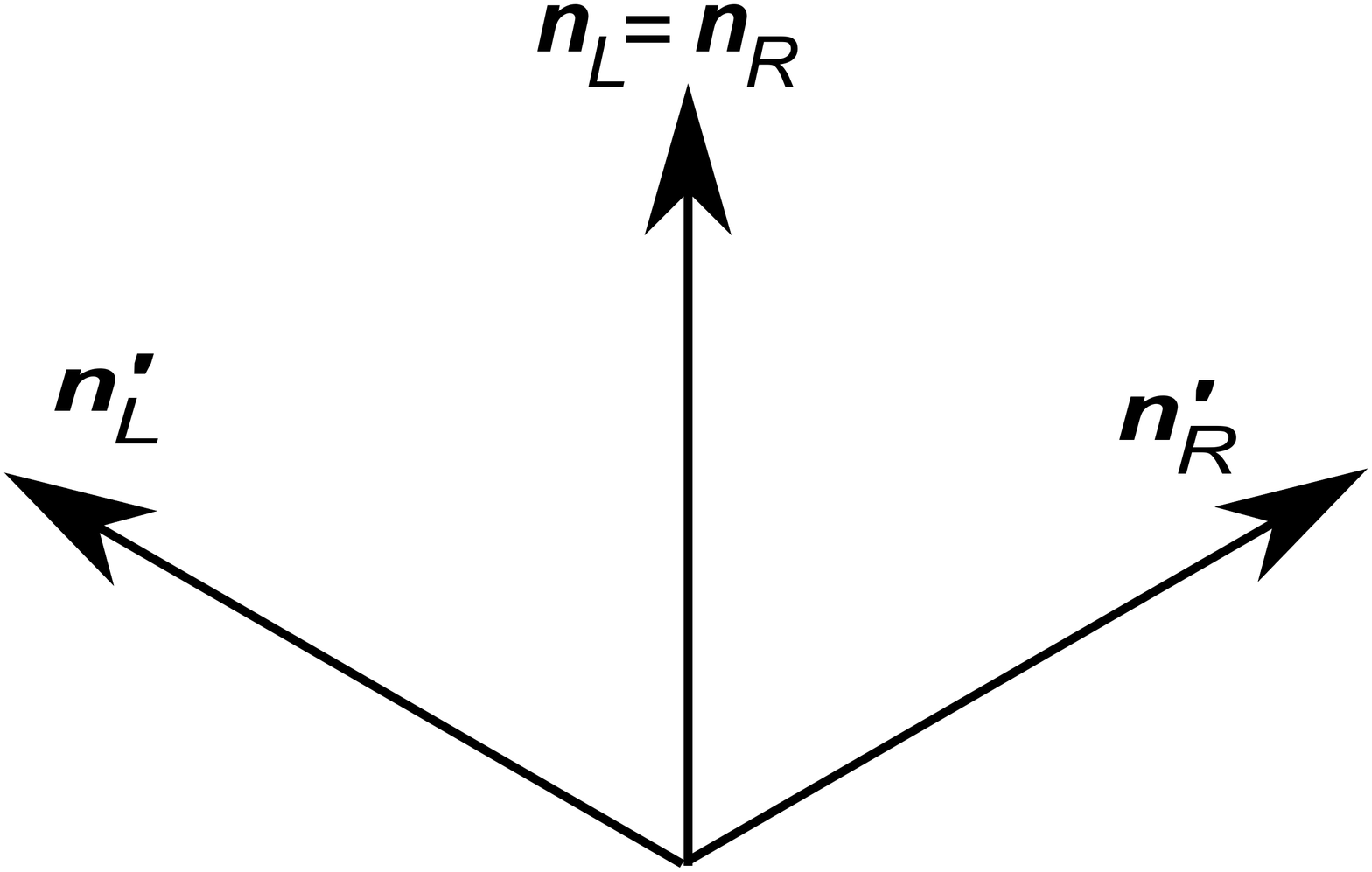}
\caption{\label{fig:maxbell} A possible configuration allowing $\mathcal{E}=4$ for the probability 
of Eq.~\eqref{eq:mixprob}.}
\end{figure}
The correlator is thus 
\begin{equation}
C(\mathbf{n}_L,\mathbf{n}_R)=-\mathrm{sgn}(\mathbf{n}_L\cdot\mathbf{n}_R) , 
\end{equation}
and the Clauser-Horne parameter 
$\mathcal{E}=|C(\mathbf{n}_L,\mathbf{n}_R)+C(\mathbf{n}'_L,\mathbf{n}_R)+C(\mathbf{n}_L,\mathbf{n}'_R)-C(\mathbf{n}'_L,\mathbf{n}'_R)|$ 
reaches the value 4 for infinitely many choices of the orientations, one of which is given in Fig.~\ref{fig:maxbell}. 

In conclusion, we have provided a realization, through classical means involving no communication 
(local or non-local), of the probabilities predicted by quantum mechanics for a spin singlet. 
While the theoretical possibility of such a decomposition was already enunciated \cite{Brans1988}, 
and two models were suggested in the recent literature \cite{Hall2010,DiLorenzo2011c}, so far 
no realization of these models was presented. Furthermore, by combining the two models above, we 
have managed to get correlations stronger than the quantum mechanical ones, so that the 
Clauser-Horne parameter reaches the maximum value, 4.

This work was supported by Funda\c{c}\~{a}o de Amparo \`{a} Pesquisa do 
Estado de Minas Gerais through Process No. APQ-02804-10.

\end{document}